\documentstyle[pra,aps,multicol,eqsecnum,epsfig]{revtex}

\newcommand\be{\begin{eqnarray}}
\newcommand\ee{\end{eqnarray}}
\newcommand\ba{\begin{array}}
\newcommand\ea{\end{array}}
\def\r{\rangle}
\def\l{\langle}
\def\T{{\rm Tr}}
\def\cP{{\cal P}}
\def\R{{\bf R}}
\def\cS{{\cal S}}
\def\cH{{\cal H}}
\def\cD{{\cal D}}

\def\cB{{\cal B}}
\def\cK{{\cal K}}
\def\cE{{\cal E}}

\def\cM{{\cal M}}

\begin{document}

\title{Quantum theory: kinematics, linearity and no-signaling condition}
\author{M\'{a}rio  Ziman${}^{1,2}$ and  Peter \v{S}telmachovi\v{c}${}^{1,2}$}
\address{
${}^{1}$ Research Center for Quantum Information,
Slovak Academy of Sciences,
D\'ubravsk\'a cesta 9, 842 28 Bratislava, Slovakia \\
${}^{2}$ Department of Mathematical Physics, National University of Ireland,
Co. Kildare, Maynooth, Ireland
}

\date{November 18th  2002}
\maketitle

\begin{abstract}
We show that the linearity of an evolution of Quantum Mechanics follows from 
 the definition of kinematics. The same result is obtained for
an arbitrary theory with the state space that includes mixtures of different
 preparations. Next, we formulate the non-signaling theorem and show 
that the theorem poses no additional restriction on Quantum Mechanics
 provided the kinematics is given. 
We also discuss validity of the postulate 
for the case of more general theories. 
\end{abstract}

\begin{multicols}{2}
\section{Introduction}
\label{sec1}

It was more than seventy  years ago when Quantum Mechanics 
became widely accepted and established as one of the fundamental 
theories of Nature. 
Despite its success there are 
still several questions which, at least for a certain 
part of the physical community, have not been answered  satisfactory yet.
The main reason is that there is much more space between  abstract 
mathematical elements of the theory and real objects prepared and
 measured  in our 
laboratory. The rigorous 
mathematical formulation of the theory was given in \cite{Neuman} which 
can be summarized in a few  postulates. 
There have always been numerous attempts to derive the postulates of
 the Quantum  Mechanics from some other more ``fundamental'' or at 
least physically well motivated postulates. As for instance to derive the 
linearity of the evolution from the no-signaling postulate \cite{Simon2001}. 
Such attempts usually raise some discussion \cite{Bona2002,Svetlichny2} 
and a casual non-expert reader may be confused by the language used by
 experts. In this paper we will try to clarify the relation 
between the linearity  and no-signaling condition in the Quantum Mechanics
 as well as present  more general results.

To begin with let us briefly summarize the most frequently  used
representation of quantum objects, i.e. the Hilbert space
formulation of quantum theory. In order to avoid certain 
mathematical complications we will
work with finite-dimensional Hilbert spaces. However, our discussion
remains valid in the case of infinite ones too.
In this framework states and observables
are associated with  specific linear operators acting on a given
Hilbert space $\cH$. States are represented by {\it density matrices}
$\varrho$, i.e. positive hermitian operators with unit trace. Let
us denote by $\cB(\cH)$ the set of all bounded linear operators
defined on the Hilbert space $\cH$. Then the subset $\cS(\cH)$
\be\label{states} \cS(\cH)=\{\varrho\in\cB(\cH)\, :\,
\varrho=\varrho^\dagger,\, \varrho\ge 0,\, \T\varrho =1\} \ee
forms a set of all possible quantum states.

Operators $O: \cH \to \cH$  for which $O=O^\dagger$ are associated with
{\it projective measurements}.  
 It turns out that a more general notion of a measurement
requires a set of operators $\{F_k\}$ for the representation of 
 a single observable $M$ \cite{Peres}. Each outcome $\lambda_k$ of the measurement $M$ is
 associated with one of these operators and the probability for
 measuring the corresponding outcome is given by the trace rule
\be
p_k = \T  \, (\varrho F_k) \;,
\ee
provided that the system was prepared in the state $\varrho$.
In the case of
projective measurements these operators possess the property
$F_j=F_j^\dagger=F_j^2$, i.e. they are {\it projective operators}.
For a general measurement $M$ these operators must be positive, 
i.e. $F_k=F_k^\dagger$ and $F_k\ge 0$,
and sum up to the identity operator, i.e. $\sum_k F_k=\openone$. Let
us denote by $\cP(\cH)$ the set of all positive elements of
$\cB(\cH)$. Usually we use the concept of operator measure defined
on Borel sets $\cB({\R})$ of real numbers $\R$ (associated
with the outcomes of the measurements). Any mapping
$F_M:\cB({\R})\to\cP(\cH)$ satisfying the properties of a measure,
i.e.
\begin{enumerate}
\item{$F_M(\R)=\openone$}
\item{$F_M(\cup_{k}A_k)=\sum_k F_M(A_k)$ where $A_k$ are mutually
disjoint Borel sets}
\end{enumerate}
represents some quantum measurement $M$. Let us denote the set
of all {\it positive operator-valued measures} (POVMs) $F_M$ with
the symbol $\cM(\cH)$.

\section{Kinematics}
When we face the problem of building some new physical theory our first step is 
to introduce basic objects representing our physical world - the kinematics. 
This can be accomplished by defining  two sets $\cS$
and $\cM$. The first one represents the states
and the second one is associated with the measurements. 
Each element  $M \in \cM$ induces a probabilistic measure $P_M$,
where
\be
\label{pm}
P_M (A, \rho)
\ee 
is the probability that the outcome of the measurement $M$ is from the Borel set $A \in \cB(\R)$ 
provided that the system was prepared in the state $\rho \in \cS$.
Here we have restricted ourselves to the $\cB (\R)$ as all measurement outcomes
are always  mapped onto the set of real numbers $\R$. An outcome of a measurement is always 
a real number like position or number of counts etc..
This probabilistic rule
is the only thing we can prove by performing our experiments. Moreover, from the mathematical
point of view, this simple rule gives us limitations on both sets in a way
that their mutual compatibility is guaranteed.
For instance, if the set of states $\cS$ is given then for any pair of states 
the set of all measurements $\cM$ must  
provide a possibility for us to distinguish the two states.
These two sets are usually called {\it kinematics} of the theory.
Now we need to associate mathematical objects with the abstract elements
of these sets, i.e. find some mathematical realization. 
In quantum theory there is a very convenient choice using the concept 
of the Hilbert space. We would like to stress here that this may  
not be the only choice, but at the same time it cannot be done in arbitrary space. 
The space should be rich enough and possess all features of the theory, like
for example interference, uncertainty relations, etc. 

Therefore, let us associate 
a Hilbert space $\cH$ with a given quantum system.
The unit elements of $\cH$ (denoted by Dirac's ket symbol $|\psi\r$),
i.e. any  vector $|\psi\r\in\cH$, for which $\l\psi|\psi\r=1$,
is an element of $\cS$ and represents a state of the quantum system.
It is well known \cite{Peres}  that the set of quantum mechanical states 
is much larger than the set composed from vector states only. 
One of the ways how to introduce  density matrices 
(mathematical objects representing
generalized quantum states) is by mixing up different pure states
 (their preparations) together.
This is how the density matrices were discovered.
``Mixing'' can be mathematically described
by {\it convex combinations} not of
the vectors, but rather as convex combinations
of  operators representing the vector states, i.e.
of {\it projectors}. More generally, mixtures can be viewed
just like probability distributions defined
on the set of unit vectors without any reference to
operators. Let us denote such set of distributions
by $\cD(\cH)$. 

{\bf Note 1:} From the operational point of view states correspond
to our preparation procedures. There are many preparations
that lead us to the same state. 
Moreover, mixing different preparations is a preparation again.
The question is whether such  mixtures prepare some new states,
or not.  In Quantum Mechanics some of these
distributions are equivalent and are represented by 
a single density matrix. This fact is
due to the given set of observables, that does not 
allow us to distinguish among different preparations of the same 
density operator. $\Diamond$

The second way how to obtain a density operator
as an object representing a state uses the notion of
composite quantum system with the Hilbert space 
given by tensor product $\cH_A\otimes\cH_B$,
where symbols $A,B$ denote two different 
physical systems. If the whole system 
is described by a vector state $|\psi\r\in\cH_A\otimes\cH_B$
then there is no guarantee that the subsystems
are described by vector states as well.
After performing partial trace operation
(discarding the second system)
we obtain that the resulting operators
possess all features of density operators.
Such confirmation gives us new possibilities how
to prepare density matrices. 
We shall call
these matrices {\it reduced density operators} and the set 
of all density matrices will be denoted as $\cS (\cH)$ 
in accordance with Eq.~(\ref{states}).
(The 
resemblance in our notation between $\cS$ and $\cS( \cH)$ is 
not accidental and will be revealed later.).

Quite natural question arises. What does happen, if we mix together
two preparations of two density matrices? Like in the case
with vector states, we can associate mixtures of density
matrices with the probability distributions
defined on the set of states $\cS(\cH)$. Let us denote 
by $\pi=\{p_j,\varrho_j\}$ such (discrete) {\it probability measure on the set of 
density operators $\cS(\cH)$} and
let $\cK(\cH)$ be the set of all such distributions $\pi$ of density matrices.
As a result we obtain new set of quantum states $\cK(\cH)$, which
 from the point of view of a mathematical description 
contains the former set $\cD(\cH)$. 
The members of $\cS(\cH)$, reduced density operators
now form extremal points of $\cK(\cH)$. 
They are not created by mixing different preparations, but by 
discarding the second system. 
Next we should find a tool that enables us to differentiate
among these states. If we find such tool (like it was find in 
the case of generalizing vector states) then we can build a  new theory 
with new set of states. We shall call any such theory 
``Extended Quantum Mechanics (EQM)'' according to Ref.~\cite{Bona2000}
 where an interesting solution to this problem has been presented. 

The set of states $\cK(\cH)$ (as well as $\cD(\cH)$) is endowed with  a convex structure. 
That is, any mixture of two probability distributions (elements of $\cK(\cH)$) is again a probability 
distribution defined on $\cS(\cH)$. 
Instead of expressing a mixing of preparations by 
$\pi=\{p_j,\varrho_j\}$ we use the natural convex structure and write any element $\pi \in \cK(\cH)$  as 
$\pi=\sum_j p_j\varrho_j$, where 
$\varrho_j$ is now a point in the state space $\cK(\cH)$ representing the state $\{ p_j=1, \varrho_j\}$.
In the formulation of the kinematics of the Quantum Mechanics  any mixture 
of density matrices, i.e an element of  $\cK(\cH)$, is associated with a certain element of 
the set $\cS(\cH)$.
That is instead of probability distributions defined 
on the set $\cS(\cH)$ we can speak only about 
the set $\cS(\cH)$ with its natural convex structure.

{\bf Note 2:} Only the convex structure of the state space $\cS(\cH)$
enables us to identify (and compare) different probability distributions 
(mixings of preparations) defined on the space $\cS(\cH)$ with the elements 
of $\cS(\cH)$. (Let us note the  vector states do not have such property;
they are just extremal points of $\cS(\cH)$).
$\Diamond$

For the time being we postpone the definition  of the set of all possible 
measurements $\cM$ for different types of state spaces. The rationale 
being that in Quantum Mechanics 
every observable is related to the generator of a one-parametric semi-group
that is a dynamical evolution. The same can be done even in the case of a more
general theory see for instance Ref. \cite{Bona2000}.

\section{Dynamics}
Under quantum dynamics two concepts can be understood.
Firstly, some rule how quantum system evolves with time,
and secondly some set of transformations of quantum states
(different from measurement) without any explicit reference to time.
Of course, any state transformation takes some time,
but we will not consider its time duration here.
This second notion determines the whole
possible set of objects in which the first rule (with time)
can draw a line to record the time dependency. In each time the state
is transformed according to a map that belongs to 
all allowable set of state transformations. Let us
now formalize these ideas. In general the allowed dynamical maps form a set
\be
\cE(\cS)=\{\Lambda\, :\, \cS(\cH)\rightarrow \cS_\Lambda\subset\cS(\cH)\}
\ee
This set contains any transformation $\Lambda$ of the quantum states.
There are yet no restrictions such as the shape of the target set $\cS_\Lambda$,
or the linearity of $\Lambda$.
In what follows by dynamics of quantum theory we will understand 
the evolution without any reference to time. It means
we will investigate the general properties,
what the dynamics should satisfy. 

Now we discuss the issue of  linearity \cite{linearity} of Quantum Mechanics.
To obtain the linearity of the evolution $\Lambda$
one has to consider the following ``mixing procedure''. 
Let us assume that there is no possibility how to distinguish  between 
two different preparations (decompositions)  of any 
density operator  by performing all possible  measurements 
$\cM(\cH)$ \cite{footnote}.
A preparator might have  this information, but in quantum theory 
(with the state space $\cS(\cH)$) we believe  it is useless. 
Let us allow nonlinear evolution $\Lambda$ and let the preparator
prepare single particle (in state $\psi_j\in\cS(\cH)$) and evolve it according to this
evolution ($\psi_j\mapsto\psi_j^\prime=\Lambda[\psi_j]$).
Such procedure is experimentally  acceptable.
Let us consider two different preparations of states $\psi_j$
and $\phi_k$ with probabilities $p_j,q_k$ such that
\begin{eqnarray*}
\sum_j p_j\psi_j=\sum_k q_k\phi_k
\end{eqnarray*}
where the equality means that the density operators are the same.
Applying the (non-linear) evolution $\Lambda$ the outgoing states
need  not represent the same density matrix, i.e.
\begin{eqnarray*}
\sum_j p_j\Lambda[\psi_j]\ne\sum_k q_k\Lambda[\phi_k]
\end{eqnarray*}
It means that the preparator is able to differentiate between
the two mixtures but then he must be able to differentiate between
them from the beginning\cite{note}. Thus we should include 
observables that enables him to do so. 
In order to preserve the kinematics (the set $\cS(\cH)$ and the set 
$\cM(\cH)$)
the non-linear evolution must be forbidden. Otherwise,
the above procedure enables us to distinguish
preparations resulting in the same density operator $\varrho$.

Let us apply  the above consideration for the Extended Quantum 
Mechanics with the  larger
set of states $\cK(\cH)$, specifically  onto the elements of $\cK(\cH)$.
If we assume that this set represents all the possible quantum states,
then the evolution defined as a mapping $\Lambda:\cK(\cH)\to\cK_\Lambda\subset\cK(\cH)$
must be ``linear'', (i.e. affine).
To show this it is enough to repeat the previous
discussion, only instead of $\cS(\cH)$ consider the set $\cK(\cH)$.
As a result we obtain that the evolution $\Lambda$ is linear in the following sense 
\begin{eqnarray*}
\Lambda[\sum_j p_j\pi_j ]=\sum_j p_j\Lambda[\pi_j]
\end{eqnarray*}
where $\pi_j\in\cK(\cH)$ and $p_j\ge 0,\ \sum_j p_j=1$.
The linearity implies that the evolution is completely determined
by its action on extremal states (Dirac distributions on $\cS(\cH)$) 
associated with the members of the set $\cS(\cH)$. 
Let us remind that $\cS(\cH)$ is not a subset of $\cK(\cH)$,
but they are different sets with different elements.
Therefore, the question of the linearity of $\Lambda$ 
on the set $\cS(\cH)$ is ill defined, because in general, 
 the map $\Lambda$ can transform
extremal states of $\cK(\cH)$ into mixtures in $\cK(\cH)$.
But if we assume that the evolution $\Lambda$ maps
extremal points into extremal points, the definition (restriction)
of $\Lambda:\cS(\cH)\to\cS(\cH)$ is possible and
the linearity can be discussed.

{\bf Note 3:} This result is nothing else but our comprehension of the notion  
probability. Probabilities arise in our description naturally
due to the preparations of mixtures.
Notice, that the set $\cK(\cH)$ is a set of all 
probability distributions defined on $\cS(\cH)$. 
It means that any element $\pi \in \cK( \cH)$ can be written as 
$\{ p_j ,\varrho_j \}$ where $\varrho_j \in \cS( \cH)$ and $\{ p_j \}$ is 
the probability distribution. Here, the elements $\pi$ are understood as 
statistical ansambles (mixings of preparations) - the state $\varrho_j$ is prepared with the 
probability $p_j$. Consequently, it must hold that
\begin{eqnarray*}
\Lambda \left[ \{ p_j, \varrho_j \} \right] = \{ p_j, \Lambda [ \varrho_j] \} \; ,
\end{eqnarray*}
since each participant from the ansamble, a system prepared in one of the  states $\varrho_j$, evolves independently on the other participants.
$\Diamond$

So far we have considered only particular cases of Quantum Mechanics
and Extended Quantum Mechanics,
but above arguments can be used for any physical theory
with specified kinematics. Therefore, we can formulate
the following theorem\\
\mbox{} \\
{\bf Theorem} {\it Consider a set of states $\cS$ and a set of measurements
$\cM$ (compatible with $\cS$), i.e. the kinematics of the theory is given. If the set 
$\cS$ is endowed with the ``convex structure'' then 
evolution must be linear.}\\ \mbox{} \\

In fact, any non-linear evolution leads either to a contradiction or to a new
kinematics. Let us suppose that the space $\cS$ is endowed with the
convex structure and an evolution $\Lambda$ is non-linear. 
The convex structure of the 
space $\cS$ is a consequence of  the possibility of 
mixing preparations. In other words for a set of elements  $\pi_j \in \cS$ 
 the mixture $\{ p_j, \pi_j \}$ is an element $\pi = \sum_j p_j \pi_j$ of
the space $\cS$ (see the end of the section Kinematics). The non-linearity of the evolution $\Lambda$ implies that 
there exist at least one pair of sets of states $\{\varrho_j\} \in \cS$ and 
$\{ \xi_i \} \in \cS$ where $\sum_j p_j \varrho_j = \sum_i q_i \xi_i$
such that
\[ \sum_j p_j \Lambda (\varrho_j) \neq \sum_i q_i \Lambda (\xi_i) \]
The two states $\sum_j p_j \varrho_j$ and $\sum_i q_i \xi_i$
represent two mixing preparations, so that
$\Lambda (\sum_j p_j \varrho_j) = \sum_j p_j \Lambda(\varrho_j)$ and 
$\Lambda (\sum_i q_i \xi_i) = \sum_i q_i \Lambda(\xi_i)$. What's more, they   
represent the same point and thus transform into a single point 
$\Lambda ( \sum_j p_j \varrho_j) = \Lambda(\sum_i q_i \xi_i)$ which is
in contradiction with the assumption that the map $\Lambda$ is non-linear as
\[ \sum_j p_j \Lambda(\varrho_j) = \Lambda (\sum_j p_j \varrho_j) = \Lambda 
(\sum_i q_i \xi_i) = \sum_i q_i \Lambda( \xi_i) \; .\]

In the case when $\cS$ is not convex 
(like for instance $\cS=\{{\rm unit\ vectors\ in}\ \cH\}$), then 
we can create a new set of states ($\cD(\cH)$), which must either be
 compatible
with the set of observables $\cM$, or we also need to change the set of 
measurements $\cM$ in order to preserve 
the mutual compatibility.
In the case of Quantum Mechanics we find out solution, where
$\cS=\cS(\cH)$ and $\cM=\cM(\cH)$. 
In particular, we can represent pure states $|\psi\r$ like
one-dimensional projections $P_\psi$ and 
define probability distributions on these projections.
The linear structure of the space of linear operators $\cB(\cH)$
enables us to associate these distributions $\{p_j,P_{\psi_j}\}$ 
with the linear operators $\sum_j p_jP_{\psi_j}\in\cB(\cH)$.
Let us remind that such assignment is a map ``from-many to-one''.
As a result of the identification of all distributions 
represented by the same operator
we obtain the set of density operators $\cS(\cH)$ which
is compatible with the set of observables $\cM(\cH)$, i.e. 
all POVMs.

{\bf Example 1:} Next we will
use the theorem to show in what
sense the evolution in Classical Mechanics
is linear, too.
{\it The phase space} $\Omega$ plays role analogical
to the Hilbert space $\cH$ (in the case of Quantum Mechanics),
or space $\cS(\cH)$ (in the case of Extended Quantum Mechanics).
It means that elements of the phase space $\Omega$ are
extremal points (denoted by $\delta_{\vec{\omega}}$) 
of the set of all classical states $\cP(\Omega)$,
i.e. {\it probability distributions} on $\Omega$.
The same arguments as before will
lead us to the linearity (on $\cP(\Omega)$) in the following sense.
Any (discrete) probability distribution
$\pi(\vec{\omega})=\sum_k\pi_k\delta_{\vec{\omega}_k}
\leftrightarrow\{\pi_k,\vec{\omega}_k\}$
must evolve with $\Lambda:\cP(\Omega)\to\cP_\Lambda\subset\cP(\Omega)$
according to the rule
\begin{eqnarray*}
\pi(\vec{\omega})=\sum_k\pi_k\delta_{\vec{\omega}_k}\mapsto
\Lambda[\pi(\vec{\omega})]=\sum_k \pi_k\Lambda[\delta_{\vec{\omega}_k}]
\end{eqnarray*}
Again, in the sense of $\cP(\Omega)$ the evolution $\Lambda$
is linear, but the transformation of the points in $\Omega$ (i.e. $\Lambda:\Omega\to\Omega$)
need not be linear.
(Unlike the Quantum Mechanics, if one starts with
the phase space $\Omega$
then the new set of classical states equals to the set of all
probability distributions $\cP(\Omega)$.)
$\Diamond$

\section{Measurements}
As has already been mentioned above the kinematics of the theory
can be viewed as a set of states $\cS$ and a set of measurements $\cM$ where
the two sets have to be mutually ``compatible''.
Our notion of  ``measurements''
correspond to ``measurable quantities'', or ``observables''. Hence,
they do not contain any description of a dynamics of the
corresponding {\it physical process of measurement}.
In the case of Quantum Mechanics the set $\cS$ is the set of all density operators $\cS(\cH)$,
while $\cM$ is the set of all positive operator
valued measures $\cM(\cH)$. And these two sets are ``compatible''. The set $\cM$ contains enough
elements so that we are able to  differentiate between any two elements of the set $\cS$.
Moreover, if we take any representation of a given state (The set $\cS$ has
a certain structure; see above) then no measurement can differentiate between
any such representations.
On the other hand if we take the set $\cS$ and ask what are all possible measurements
i.e. all possible probability measures defined on the set $\cS$ then 
we find out that the sought set is $\cM$.

In order to retain the larger set of states $\cK(\cH)$ of Extended Quantum
Mechanics, but still use only density operators (i.e. elements of $\cS(\cH)$)
 for our description of states, we must be able to  differentiate between two different
 types of density matrices: {\it genuine mixtures} and {\it elementary mixtures} (for more details see section Kinematics or Ref.~\cite{Bona2000}). 
The elementary mixtures represent reduced density operators
and genuine mixtures are associated with statistical mixtures of these reduced
density operators. However, in order to discriminate between an elementary mixture (reduced density operator) 
$\varrho$ and a genuine mixture 
$\{\lambda_j,\omega_j\}$ 
(with $\omega_j \in \cS(\cH)$ being elementary mixtures) associated with
the same density operator $\varrho=\sum_j\lambda_j\omega_j$ (decomposition is fixed),
we have to introduce an observable $M$ that is  non-linear (see Eq.~(\ref{pm}))
\begin{equation}
P_M(A, \sum_j \lambda_j \omega_j) \neq \sum_j \lambda_j P_M(A,\omega_j) \; ,
\end{equation}
for at least one $A \in \cB(\R)$. Consequently, if we decide to deal with
the set of density operators $\cS(\cH)$,  then the set $\cM$ 
has to include non-linear observables.
Let us remind the reader that quantum mechanical observables, 
self adjoint operators, are in Quantum Mechanics
identified with the generators of a dynamical evolution. Once we allow
non-linear evolution then we have to include observables that are non-linear and vice versa. But let us stress here that the evolution of genuine mixtures (elements of $\cK(\cH)$) is always linear
while elementary mixtures (elements of $\cS(\cH)$) can evolve according
to non-linear maps.

\section{No-signaling condition}
The impossibility to transmit information faster than the  propagation 
of light seems to be an interesting problem in the context of quantum theory. 
Undoubtedly, in quantum theory there is no dynamical 
restriction how fast the particles can move.
However, no-signaling does not deal with dynamical properties
of the theory, but rather kinematic ones, namely with the possibility
to use the {\it projection postulate} of quantum measurements
in the information transfer \cite{Comment}. Naturally, only the theory with 
the projection postulate (or more generally
any postulate of similar type) can be questioned. 
Therefore, our following discussion will be focused
only on two specific theories: Quantum
Mechanics and Extended Quantum Mechanics (for definition
see sections Kinematics and Measurements).

Let us consider a projective measurement $M$ represented by the set 
of projective operators $\{ F_k\}$ 
(for more details see Introduction and Ref.~\cite{Peres})
and let a quantum  system be prepared in a state $\varrho$.
The projection postulate states:\\ \mbox{} \\
{\it 
After performing  a projective measurement $M$  
(in non-demolition experiments) resulting in the observation
of the value $\lambda_k$ the system is described by the following state
\be
\label{projection}
\varrho_k=\frac{F_k\varrho F_k}{\T\varrho F_k} \; .
\ee }

The postulate is an independent postulate of Quantum Mechanics
and has been introduced on account of the following:
When we repeat the same measurement (on the same object) 
right after the first one 
then the results of the two consecutive measurements 
are always the same. 
After many repetitions of the same measurement 
on systems all prepared  in the state $\varrho$ 
(i.e. following the same preparation process)
the final ensemble will be described
by a mixture of states $\varrho_k$ associated with different outcomes, i.e.
$\varrho_f=\{p_k,\varrho_k\}=\sum_k p_k\varrho_k$ 
with $p_k=\T\varrho F_k$. As a result we have that measurements (without postselection)
prepare systems in mixtures. That is, for anybody who does not have
access to observed values, the outcome of the measurement is described
by this mixture $\varrho_f$. Moreover, another measurement $\widetilde{M}$ (with 
corresponding operators $\{\widetilde{F}_\alpha\}$) 
can result in a different mixture $\{\widetilde{p}_\alpha, \widetilde{\varrho}_\alpha \}$.

{\bf Example 2:} 
Let us consider a pair of two-dimensional quantum systems (qubits)
 denoted as  $A$ and $B$ prepared in the state
\begin{eqnarray*}
|\psi\r=\frac{1}{\sqrt{2}} \left \{ |0\r_A\otimes|1\r_B-|1\r_A\otimes |0\r_B \right \}  \, .
\end{eqnarray*}
To demonstrate the projection postulate  we will consider  a specific 
measurement $M$ represented by two operators $F_k$ defined as 
\begin{eqnarray*}
F_0 & = & \openone_A \otimes |0 \rangle_B \langle 0| \\
F_1 & = & \openone_A \otimes |1 \rangle_B \langle 1| \; .
\end{eqnarray*}
It is easy to see that the measurement $M$ is actually a projective 
 measurement acting  on the system $B$
only with two  projectors  $|0 \rangle_B \langle 0|$ and $| 1 \rangle_B \langle 1 |$. 
After obtaining the first outcome the bipartite system is according to Eq.~(\ref{projection}) 
in the state
$ \varrho_0= |1\r_A \l 1 | \otimes|0 \r_B \l 0 |$ 
and when the second outcome is measured the system is in the state
$ \varrho_1 = |0\r_A \l 0 | \otimes|1 \r_B \l 1 | $.
Thus, by measuring the bipartite system we prepare a 
mixture of states $\varrho_0$ and $\varrho_1$ with equal probabilities as
$\T |\psi \rangle \langle \psi | F_0 = \T |\psi \rangle \langle \psi | F_1 = 1/2$. 
What is more interesting is the fact that by measuring the subsystem $B$ the state 
of the subsystem $A$ changes as well. For instance, if the eigenvalue $\lambda_0$ 
is measured then the state of the bipartite system is $\varrho_0$ and the state 
of the subsystem $A$ is
\begin{eqnarray*}
\T_B \varrho_0 = | 1 \rangle_A \langle 1 |
\end{eqnarray*}
which is different from its original state
\begin{eqnarray*}
\rho_A = \T_B | \psi \rangle \langle \psi| = \frac{1}{2} \openone_A
\end{eqnarray*}
$\Diamond$

The fact that a measurement performed 
on the system $B$ can change a description
of the state of the system $A$ no matter how far from each other they are
is certainly a peculiar property. What is more this change is considered to be 
instantaneous. Therefore, it is correct to ask, whether such  property 
cannot be used for signaling or transmission of information at speeds 
larger than the speed of light. 

In general, consider a bipartite system where $A$ and $B$ 
are corresponding parts and let the system be in a state $\varrho$. 
Any projective measurement $M$
performed on the system $B$  can be represented 
with operators $F_k = \openone_A \otimes P_k$,
where $P_k$ are operators (projectors) acting on the system $B$ only. 
After measuring $M$ the state of the system $A$ is (according 
to Eq.~(\ref{projection})) in one of the states $\T_B \varrho_k$ 
with probability $p_k$.
Due to the fact that the only thing we can predict are the probabilities
 of individual outcomes of a given 
measurement but not which of them  is observed in a single event, 
it follows that by measuring the system $B$ we prepare the system $A$ in the mixture
$\varrho_A^M = \{p_k, \T_B \varrho_k \}$. (Let us note here that the observer 
possessing the system $A$ does not know the results of the measurements 
performed on the system $B$.)
For different measurements $M$ the mixture
$\varrho_A^M$ can be different \cite{Comment2}. 
However, if we express the  $\varrho_A^M$ as a 
density operator and use Eq.~(\ref{projection})  then it is easy to see that 
\begin{equation}
\label{decomp}
\varrho_A^M = \sum_k p_k \T_B \varrho_k  = \T_B \varrho = \varrho_A \, .
\end{equation} 
It means, that the resulting state of the system $A$ 
is described by the original density operator $\varrho_A$,
only its decomposition is different.

As we have shown,
by using different measurements on the first 
part of a bipartite system we can prepare different realizations (mixtures) of a given 
density operator of the second part. 
It means that the only information we can ``signal'' using 
this procedure is the information on the particular realization 
of a given density operator. 
But the {\em kinematics} of the Quantum Mechanics is such that two different 
realizations of a given density operator represent the same state 
(see the section Kinematics) and  there is no measurement that
the owner of the second  system  could use to distinguish the two preparations.  
In other words, within the standard quantum state space $\cS(\cH)$
two different realization of $\varrho_A$ (two different statistical
mixtures or a statistical mixture and the reduced density operator) represent the
same point. Therefore, the no-signaling holds and follows from the kinematic
properties of the set $\cS(\cH)$.

In the context of the Extended Quantum Mechanics 
with the state space $\cK(\cH)$ 
the situation is different.
The projection postulate 
corresponds to the projection onto a
Dirac distribution $\delta_\psi=\{1,P_\psi\}$, where $P_\psi$ 
is a projector associated with the vector state $|\psi\r\in\cH$. 
If we apply this postulate onto bi-partite systems,
then we are able to prepare two different elements of $\cK(\cH)$ 
from a spatially distant place in the Universe (see Eq.~(\ref{decomp}) 
and discussion above). Let us remind the reader that in the theory 
with the state space 
$\cK(\cH)$ two different decompositions of a given density operator 
represent different points in $\cK(\cH)$. 
Therefore, the situation is different from that in Quantum Mechanics. 
Due to the experimental possibility to discriminate
two decompositions of a given density operator, our information transfer
(based on this property) will be as fast as we are
able to distinguish the two prepared states (the projection is
considered to be instantaneous).
We can always place the second system far enough from the first one to violate
the second principle of relativity, i.e. we will be able to signal 
at a  speed greater than the speed of light.

In conclusion, in the Extended Quantum Theory 
with the projection postulate
the no-signaling condition does not hold \cite{last}.
On the other hand, in Quantum Mechanics
the no-signaling holds and follows from the kinematics
 of Quantum Mechanics. We have also shown 
(see section Dynamics) that the linearity follows 
from the kinematic properties too. Let us stress here 
that this result is independent on the no-signaling condition  which,
 in the particular case of kinematics of Quantum Mechanics, 
is therefore redundant. 
Finally, even though the Extended Quantum Mechanics
together with the projection postulate is not compatible with
the no-signaling condition we cannot exclude all non-linear theories
(for instance those not using the projection postulate),
 inasmuch as we have not considered 
the most general case.

\acknowledgements
We would like to thank Pavol B\'ona  
for useful comments and Vladim\'{\i}r Bu\v{z}ek for discussion.
This was work supported in part
by  the European Union projects EQUIP (IST-1999-11053) and
QUBITS (IST-1999-13021).

\end{multicols}

\begin{thebibliography}{99}

\bibitem{Neuman} J. von Neumann, 
{\em Mathematische Grundlagen der Quantenmechanik}, Springer, Berlin (1932)
\bibitem{Simon2001} Ch. Simon, N.Gisin, V.Bu\v zek, 
{\em Phys. Rev. Lett.} {\bf 87} 170405 (2001)
\bibitem{Bona2002} P. B\'ona, Commnet on ``No Signaling Condition and Quantum Dynamics'', quant-ph/0201002.
\bibitem{Svetlichny2} George Svetlichny, Critique of ``No Signaling Condition and Quantum Dynamics'', quant-ph/0208049.
\bibitem{Peres} see for instance A. Peres, {\em Quantum Theory: Concepts and Methods}, Kluwer Acadenic Publishers, Dordrecht.
\bibitem{Bona2000}P. B\'ona, {\it Acta Phys. Slovaca} {\bf 50}, 1-198 (2000)
\bibitem{linearity}
The notion of the linearity of an evolution
$\Lambda$ makes sense, only when the set $\cS$
is a subset of the linear (vector) space. In that case we shall call 
a transformation $\Lambda:\cS\to\cS$ linear,  
if for all $\lambda_j$ and all $\omega_j\in\cS$ 
(such that $\sum_j\lambda_j\omega_j\in\cS$) holds
$\Lambda[\sum_j\lambda_j\omega_j]=\sum_j \lambda_j\Lambda[\omega_j]$.
Otherwise the evolution $\Lambda$ is non-linear.
\bibitem{footnote}
 This assumption determines  the state space of quantum 
theory to be $\cS(\cH)$. For more details see section Kinematics.
\bibitem{note} In principle, we might consider evolutions 
where a single state evolves into two or more different states. 
There is a deep reason why not to allow any evolution of that sort. 
The way physics has been developed  is that 
given the initial conditions and all interactions, the
evolution (not measurement) of the system is deterministic, i.e.
(``dynamical'') future of the system is completely predicted
by its initial state and given dynamics.
If this were not true
then we would need to change not only our theory but also our
way of looking at the world. 
However, this has nothing to do with the possible distinguishability
of different preparations of a  density operator.
\bibitem {Comment} From this point of view it makes no sense to 
speak about the no-signaling without the projection postulate as was done in the introduction in \cite{Simon2001}. But we also  should  say that despite the authors claims in the indroduction  they later used it in subsequent paragraphs of the cited paper.
\bibitem{Comment2}
This holds for any
non-factorizable state $\varrho_{AB}\ne\varrho_A\otimes\varrho_B$ 
where $\rho_{AB} \in \cS(\cH)$.
\bibitem{last}
To avoid the non-signaling in the variant of quantum theory with
larger state space $\cK(\cH)$, perhaps we should reconsider
the validity of the projection postulate.
We should exclude instantaneous action onto the second system. Probably we will need 
to introduce some dependency   
on the mutual distance of the measuring device and the affected system.
But we do not want to force any new idea about this reconsideration.

\end{thebibliography}
\end{document}